\documentclass[runningheads]{llncs}
\usepackage{marvosym} %
\usepackage[T1]{fontenc}
\usepackage{amsmath}
\usepackage{amssymb}
\usepackage{latexsym}
\usepackage{graphicx,verbatim}
\usepackage{hyperref}
\usepackage{makecell}
\usepackage{multirow}
\usepackage{makecell}
\usepackage{bm}

\usepackage{xcolor}
\def\xblue{\textcolor{blue}}
\begin{document}
%
\title{Unleashing Vision Foundation Models for Coronary Artery Segmentation: Parallel ViT-CNN Encoding and Variational Fusion}
\titlerunning{Unleashing VFMs for Coronary Artery Segmentation}
%

\author{
	Caixia Dong\inst{1,2} \and 
	Duwei Dai\inst{2} \and 
	Xinyi Han\inst{3} \and
	Fan Liu\inst{2} \and 
	Xu Yang\inst{2} \and 
	Zongfang Li\inst{1,2}\textsuperscript{(\Letter)} \and
	Songhua Xu\inst{1,2}\textsuperscript{(\Letter)}
}

\authorrunning{C. Dong et al.}

\institute{
	National-Local Joint Engineering Research Center of Biodiagnosis $\&$ Biotherapy, the Second Affiliated Hospital of Xi’an Jiaotong University, Xi’an, 710004, China \\
	\and
	Institute of Medical Artificial Intelligence, the Second Affiliated Hospital of Xi’an Jiaotong University, Xi'an, 710004, China \\
	\email{lzf2568@xjtu.edu.cn, songhuaxu@mail.xjtu.edu.cn}
	\and
	Viadrina European University, Frankfurt (Oder), 15230, Germany  \\
}
   
\maketitle              

\begin{abstract}
Accurate coronary artery segmentation is critical for computer-aided diagnosis of coronary artery disease (CAD), yet it remains challenging due to the small size, complex morphology, and low contrast with surrounding tissues. 
To address these challenges, we propose a novel segmentation framework that leverages the power of vision foundation models (VFMs) through a parallel encoding architecture. Specifically, a vision transformer (ViT) encoder within the VFM captures global structural features, enhanced by the activation of the final two ViT blocks and the integration of an attention-guided enhancement (AGE) module, while a convolutional neural network (CNN) encoder extracts local details. These complementary features are adaptively fused using a cross-branch variational fusion (CVF) module, which models latent distributions and applies variational attention to assign modality-specific weights. Additionally, we introduce an evidential-learning uncertainty refinement (EUR) module, which quantifies uncertainty using evidence theory and refines uncertain regions by incorporating multi-scale feature aggregation and attention mechanisms, further enhancing segmentation accuracy. 
Extensive evaluations on one in-house and two public datasets demonstrate that the proposed framework significantly outperforms state-of-the-art methods, achieving superior performance in accurate coronary artery segmentation and showcasing strong generalization across multiple datasets.
The code is available at \href{https://github.com/d1c2x3/CAseg}{https://github.com/d1c2x3/CAseg}.


\end{abstract}
	\begin{figure*}[!ht]
	\centering
	\includegraphics[width=0.999\linewidth]{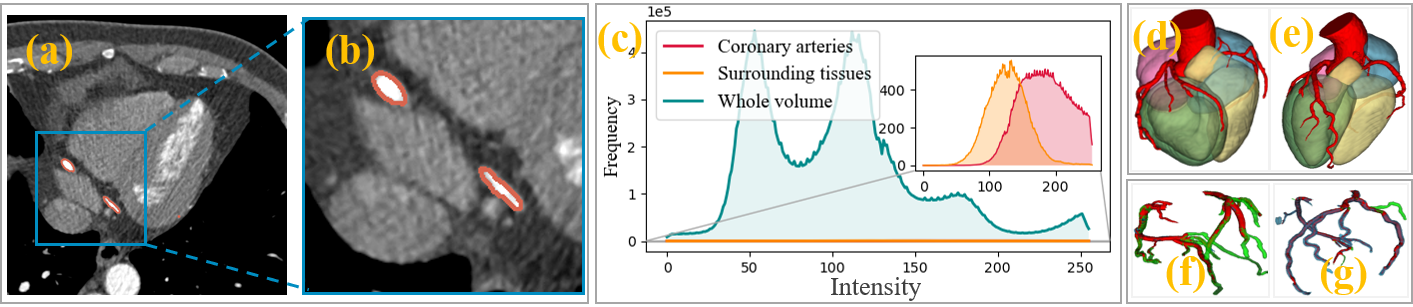}
	\caption{The major challenges of accurate coronary artery segmentation. In (f) and (g), red, green, and cyan indicate correct, under, and over-segmentation, respectively.
	}
	\label{challenges}
	\vspace{-6ex}
\end{figure*}
\section{Introduction}
Coronary artery disease (CAD) is the most common type of heart disease and a leading cause of global mortality \cite{virani2021heart}. Given the significant clinical challenges associated with CAD, accurate imaging techniques are critical for early diagnosis and effective treatment planning. Coronary computed tomography angiography (CCTA) has become as the standard non-invasive modality for evaluating coronary artery anatomy and pathologies \cite{marano2020ccta}. Precise segmentation of coronary arteries in CCTA images is essential for assessing stenosis severity, plaque morphology, and guiding clinical decision-making in CAD management.

Despite advancements in imaging technology, accurate coronary artery segmentation in CCTA images remains challenging due to several inherent factors: small vessel size (Fig. \ref{challenges}(a)-(b)), low contrast with surrounding tissues (Fig. \ref{challenges}(c)), and complex vascular morphology (Fig. \ref{challenges}(d)-(e)), all of which complicate the task of delineating vascular structures.

Deep learning has shown significant potential in coronary artery segmentation, offering improved scalability and accuracy. UNet and its variants remain foundational to many state-of-the-art models \cite{song2022automatic,dong2025high,dong2023novel}. For example, 3D-FFR-UNet \cite{song2022automatic} improves feature fusion with dense convolutional blocks, while Dong et al. \cite{dong2023novel} leverage multi-scale attention to capture finer vessel details. However, while these convolutional neural network (CNN)-based methods are effective at extracting local features, they often struggle to preserve the anatomical continuity of vessels, resulting in fragmented and anatomically inconsistent segmentations, particularly in complex vascular regions (Fig. \ref{challenges}(f)).
Vision transformer (ViT)-based approaches \cite{zhou2023nnformer,hatamizadeh2022unetr}, in contrast, excel at modeling global structural features but often lack the spatial resolution needed to preserve fine-grained vascular details essential for delineating thin and tortuous vessels (Fig. \ref{challenges}(g)).
 Hybrid approaches that combine CNNs and ViTs offer promising solutions. For instance,  Pan et al. \cite{pan2022deep} propose a cross-transformer network that integrates UNet for local features and Transformers for long-range dependencies. Similarly, Ensembled-SAMs \cite{chen2024ensembled} integrates nnU-Net \cite{isensee2021nnu} with SAMs \cite{kirillov2023segment} but rely on 2D slice processing and result merging, neglecting feature-level fusion and 3D inter-slice continuity.
 
\begin{figure*}[!t]
		\centering
		\includegraphics[width=0.9999\linewidth]{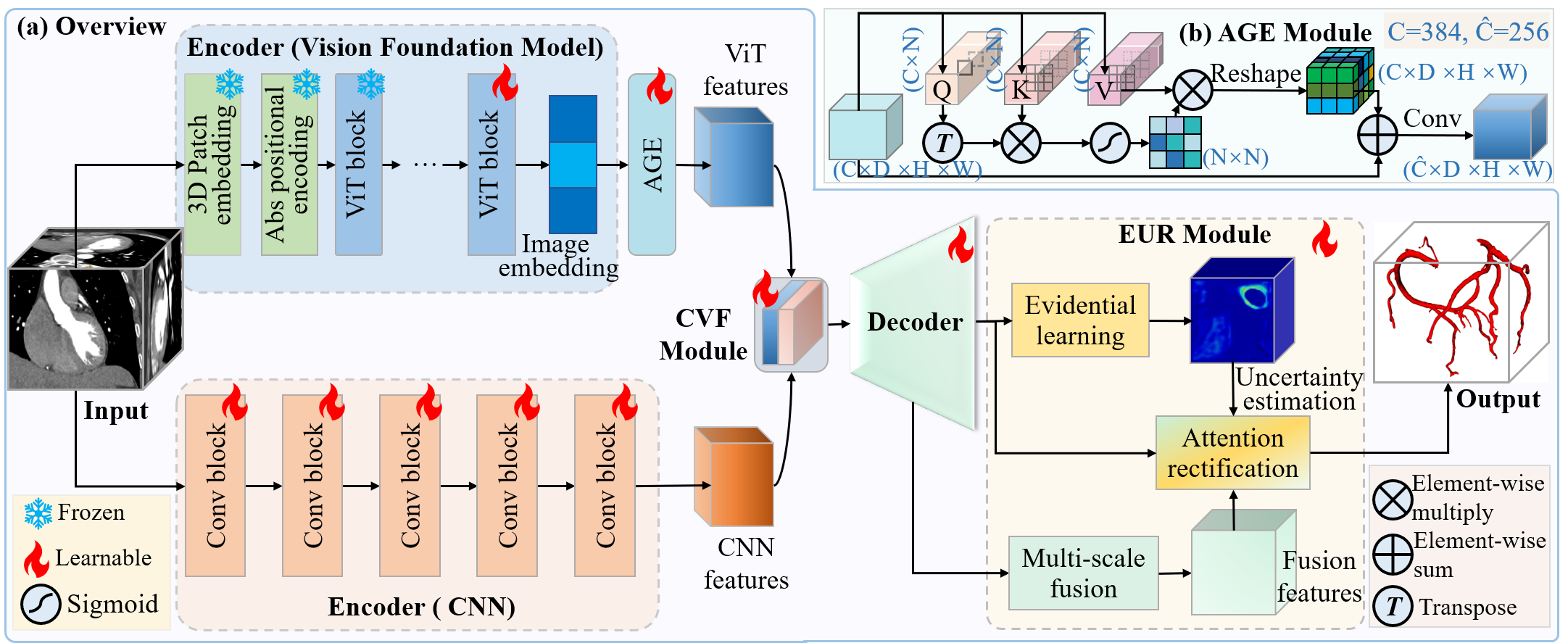}
		\caption{Illustrates the pipeline of the proposed framework. It consists of three main components. (1) A parallel encoding architecture integrates a 3D foundation model-driven encoder and a 3D UNet-based shape encoder to extract  3D volumetric representations from global and local perspectives. (2) A CVF module adaptively fuses these features by modeling latent distributions and applying variational attention. (3) An EUR module refines predictions in uncertain regions to enhance segmentation accuracy.
		}
		\label{frame}
		 \vspace{-2ex}
\end{figure*}

 In this work, we propose a novel segmentation framework that leverages the power of vision foundation models (VFMs) through a parallel encoding architecture (Fig. \ref{frame}). 
 \textbf{First}, the ViT encoder within the VFM \cite{wang2023sam} captures global structural features, enhanced by the activation of the final two ViT blocks and the integration of an attention-guided enhancement (AGE) module, which improves the model's ability to capture vascular continuity and topology; meanwhile, the CNN encoder extracts local details, ensuring a comprehensive representation. \textbf{Second}, to effectively fuse global and local information, we introduce a cross-branch variational fusion (CVF) module, which models latent distributions and applies a variational attention mechanism to adaptively assign modality-specific weights. \textbf{Additionally}, we design the evidential-learning uncertainty refinement (EUR) module to quantify segmentation uncertainty using evidence theory and refine predictions by aggregating multi-scale features and attention mechanisms. 
 These components collectively enhance segmentation accuracy and robustness, particularly in complex vascular structures.

	\section{Methodology}
\label{sec3}
An overview of the proposed framework is shown in Fig. \ref{frame}(a). It employs a parallel encoding architecture, combining a 3D foundation model-driven encoder (e.g., based on a pre-trained like SAM-Med3D \cite{wang2023sam}) with a 3D UNet-based shape encoder. These two encoders work in parallel to extract complementary 3D volumetric representations, capturing global contextual information and local structural details simultaneously. To enhance global feature perception, the final two ViT blocks are activated, and an AGE module is integrated  (Fig. \ref{frame}(b)), which leverages attention mechanisms \cite{dong2023novel} and a fusion
layer to emphasize vascular continuity and morphology. Next, we introduce a CVF module to fuse these features obtained from the two encoders. Finally, an EUR module is introduced to refine predictions in uncertain regions.

The details of our method are elaborated in the following sections.

\subsection{Cross-branch Variational Fusion Module}
The CVF module is designed to integrate global and local features extracted from the ViT and CNN branches. 
The module comprises two core components: latent distribution learning and variational attention fusion.
\begin{figure*}
	\centering
	\includegraphics[width=0.85\linewidth]{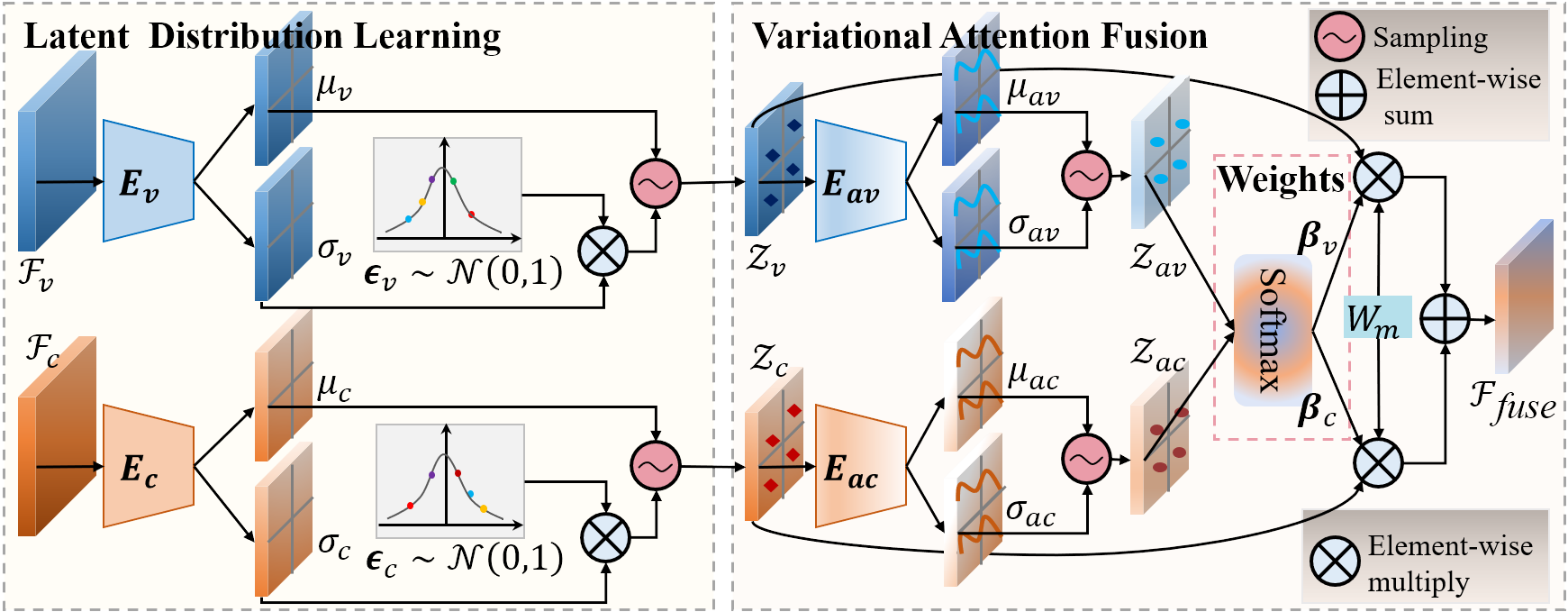}	
	\caption{Structure of the cross-branch variational fusion module. It integrates global and local features through latent distribution learning and variational attention fusion.Two encoders \(E_v\) and \(E_c\) parameterize the mean and variance of latent distributions, while \(E_{av}\) and \(E_{ac}\) compute adaptive weights \(\beta_v\) and \(\beta_c\) for feature fusion. 
	}
	\label{fig:cvf}
	 \vspace{-2ex}
\end{figure*}

\textbf{Latent Distribution Learning.}
The CVF module employs independent encoders, \(E_v\) and \(E_c\), for the ViT and CNN branches to capture the inherent variability and complementarity of global and local features. These encoders utilize multi-layer perceptrons (MLPs) to parameterize the latent distributions of global (\(\mathbf{F}_v\)) and local (\(\mathbf{F}_c\)) features, modeling them as Gaussian distributions with learnable means and standard deviations:
\begin{equation}
	\boldsymbol{\mu}_v = \mathrm{MLP}(\mathbf{F}_v),\quad
	\boldsymbol{\sigma}_v = \mathrm{MLP}(\mathbf{F}_v),\quad
	\boldsymbol{\mu}_c = \mathrm{MLP}(\mathbf{F}_c),\quad
	\boldsymbol{\sigma}_c = \mathrm{MLP}(\mathbf{F}_c).
\end{equation}
The latent variables \(\mathbf{Z}_v\) and \(\mathbf{Z}_c\) are then sampled as \(\mathbf{Z}_v \sim \mathcal{N}(\boldsymbol{\mu}_v, \boldsymbol{\sigma}_v^2)\) and \(\mathbf{Z}_c \sim \mathcal{N}(\boldsymbol{\mu}_c, \boldsymbol{\sigma}_c^2)\).
To ensure differentiability during training, the reparameterization trick is applied:
$\mathbf{Z}_v = \boldsymbol{\mu}_v + \boldsymbol{\sigma}_v \cdot \boldsymbol{\epsilon}_v$ and $\mathbf{Z}_c = \boldsymbol{\mu}_c + \boldsymbol{\sigma}_c \cdot \boldsymbol{\epsilon}_c$, where $\boldsymbol{\epsilon}_v, \boldsymbol{\epsilon}_c \sim \mathcal{N}(\mathbf{0}, \mathbf{I})$.

This mechanism enables the CVF module to learn robust feature representations that account for both deterministic and stochastic variations. Consequently, the latent variables encapsulate richer contextual information, which is critical for downstream tasks.

\textbf{Variational Attention Fusion.}
The global and local latent features are integrated using a variational attention mechanism. 
First, the latent variables \(\mathbf{Z}_v\) and \(\mathbf{Z}_c\) are processed through MLP-based encoders \(E_{av}\) and \(E_{ac}\), generating intermediate latent distributions: \(\mathbf{Z}_{av} \sim \mathcal{N}(\boldsymbol{\mu}_{av}, \boldsymbol{\sigma}_{av}^2)\) and \(\mathbf{Z}_{ac} \sim \mathcal{N}(\boldsymbol{\mu}_{ac}, \boldsymbol{\sigma}_{ac}^2)\). Similar to \(E_v\) and \(E_c\), these encoders ensure consistent and robust latent feature representation. 
Fusion weights \((\boldsymbol{\beta}_v, \boldsymbol{\beta}_c)\) are then computed via the softmax function: \((\boldsymbol{\beta}_v, \boldsymbol{\beta}_c) = \mathrm{Softmax}(\mathbf{Z}_{av}, \mathbf{Z}_{ac})\). The final fused feature representation \(\mathbf{F}_{\mathrm{fuse}}\) is computed as a weighted combination of the latent variables from both branches:
\begin{equation}
	\mathbf{F}_{\mathrm{fuse}} = \mathbf{W}_m \cdot (\boldsymbol{\beta}_v \cdot \mathbf{Z}_v + \boldsymbol{\beta}_c \cdot \mathbf{Z}_c),
\end{equation}
where \(\mathbf{W}_m\) is a learnable weight matrix for optimal feature transformation.

This mechanism adaptively balances global and local contributions, enhancing the model's ability to capture both macro- and micro-level vessel structures.

\subsection{Evidential-learning Uncertainty Refinement Module}

\label{sec3-3}
The EUR module enhances segmentation robustness in ambiguous and low-contrast regions through evidential uncertainty estimation, multi-scale feature fusion, and uncertainty-guided refinement. 
\begin{figure*}[!t]
	\centering
	\includegraphics[width=0.999\linewidth]{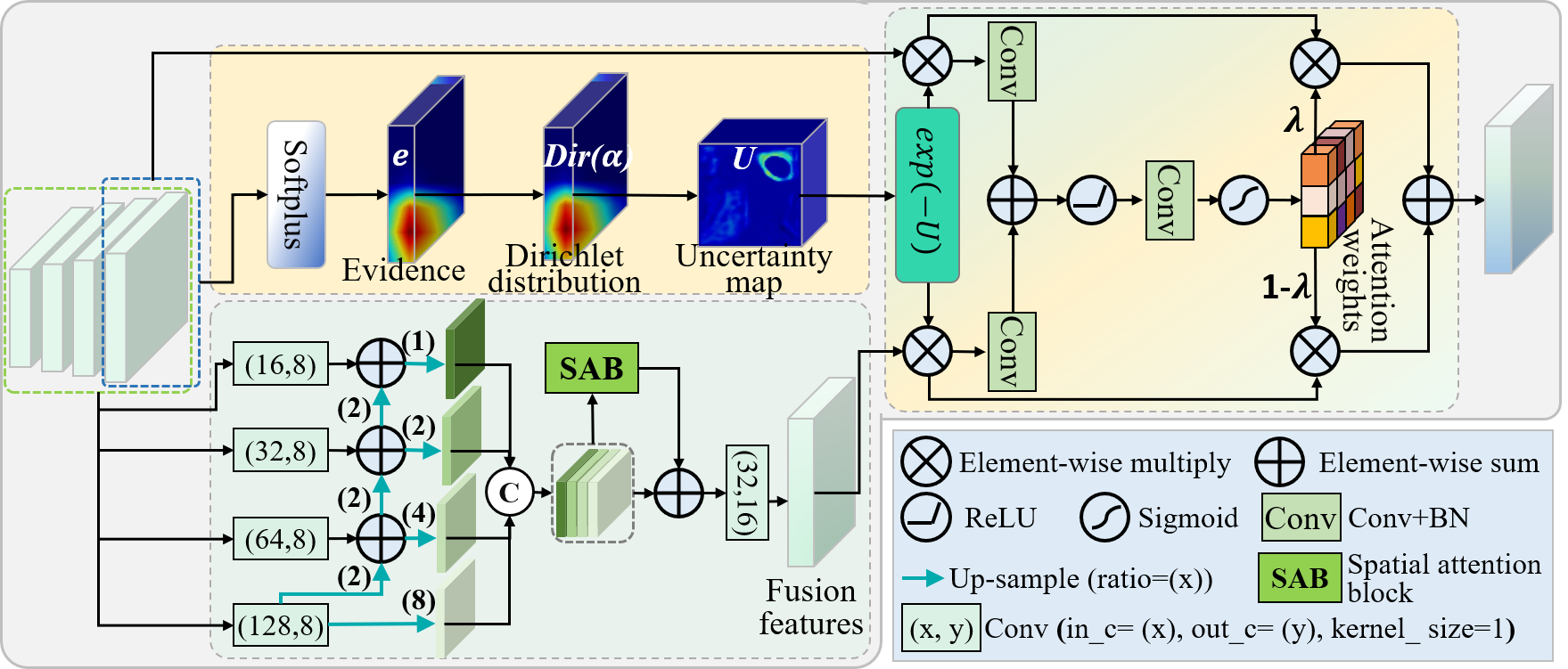}
	\caption{Structure of the evidential-learning uncertainty refinement module. The module refines segmentation by combining evidential uncertainty modeling, multi-scale feature fusion, and uncertainty-guided refinement.
	}
	\label{fig:eur}
	 \vspace{-5ex}
\end{figure*}

\textbf{Uncertainty Quantification.}
Deep models often exhibit overconfidence, reducing reliability in medical segmentation. To address this, the EUR module employs evidential learning paradigm based on Subjective Logic theory \cite{jsang2018subjective}, modeling uncertainty through evidence rather than direct probabilities.

A Dirichlet distribution \cite{xu2024eviprompt} is adopted to capture voxel-wise uncertainty, with the evidence map \(\mathbf{e}\) computed via a non-negative activation function, Softplus: \(\mathbf{e} = \mathrm{Softplus}(\mathbf{F})\). Here \(\mathbf{F}\) represents the input feature map. The Dirichlet parameters are then given by \(\boldsymbol{\alpha} = \mathbf{e} + 1\), where \(\boldsymbol{\alpha} = [\alpha_1, \ldots, \alpha_K]\), and \(K\) is the number of classes. Uncertainty is estimated as:
$U = \frac{K}{S}$, where $S = \sum_{k=1}^K \alpha_k$ denotes the Dirichlet strength.
This formulation highlights high-uncertainty regions such as boundaries and low-contrast areas, guiding more informed segmentation.

\textbf{Multi-scale Feature Fusion.}
To enhance the network's ability to capture contextual information, the EUR module integrates multi-scale features from different decoder stages. Lower-resolution features are upsampled to align with higher-resolution ones, followed by progressive fusion:
\begin{equation}
	\mathbf{F}_i^{t} = \left\{\begin{array}{ll}
		\mathrm{Conv}(\mathbf{F}_i), & i = 4 \\
		\mathrm{Conv}(\mathbf{F}_i) + \mathrm{Up}_{\times 2}(\mathbf{F}_{i+1}^{t}), & i = 3, 2, 1
	\end{array}\right.
	, \quad
	\mathbf{F}_i^{\prime} = \mathrm{Up}_{\times 2^{i-1}}(\mathbf{F}_i^{t}),
\end{equation}
where \(\mathbf{F}_i\) denotes features from the \(i\)-th scale, \(\mathrm{Conv}(\cdot)\) denotes a 1×1×1 convolution for channel alignment, and \(\mathrm{Up}_{\times k}(\cdot)\) denotes upsampling with a ratio of \(k\). The fused features are concatenated as \(\mathbf{F}_c = \mathrm{Cat}(\mathbf{F}_1^{\prime}, \mathbf{F}_2^{\prime}, \mathbf{F}_3^{\prime}, \mathbf{F}_4^{\prime})\), where \(\mathrm{Cat}(\cdot)\) denotes the concatenation operation. A spatial attention block (SAB) \cite{liao2022real} further enhances spatial localization:
$
\mathbf{F}_{\mathrm{fusion}} = \mathbf{F}_c + \mathrm{SAB}(\mathbf{F}_c).
$
This fusion strategy enhances cross-scale interaction and spatial sensitivity, ensuring robust feature representations.

\textbf{Uncertainty-guided Refinement.}
The EUR module refines the segmentation results by integrating the initial prediction \(\mathbf{P}\), uncertainty map \(\mathbf{U}\), and fused features \(\mathbf{F}_{\mathrm{fusion}}\). First, a reliable mask \(\mathbf{M}_r\) is constructed to suppress uncertain regions:
$
\mathbf{M}_r = (\mathbf{P} + \mathbf{F}_{\mathrm{fusion}}) \cdot \exp(-\mathbf{U}).
$
Here, \(\exp(-\mathbf{U})\) suppresses high-uncertainty areas, focusing on more reliable regions. Next, an attention mechanism \cite{oktay2018attention} adaptively highlights important spatial regions by generating a dynamic weight map \(\boldsymbol{\lambda} = \mathrm{Sigmoid}(\mathrm{Conv}(\mathrm{ReLU}(\mathbf{M}_r)))\), where \(\boldsymbol{\lambda} \in [0, 1]\). The final refined representation is obtained as:
\(
\mathbf{F}_{\mathrm{refined}} = \boldsymbol{\lambda} \cdot \mathbf{P} + (1-\boldsymbol{\lambda}) \cdot \mathbf{F}_{\mathrm{fusion}}
\),
where \(\boldsymbol{\lambda}\) balances the contributions of the initial prediction and fused features to improve accuracy.
\subsection{Loss Function}

Our training objective integrates a combined segmentation loss (\(\mathcal{L}_{\mathrm{seg}}\) \cite{dong2024novel}) and an evidential regularization loss (\(\mathcal{L}_{\mathrm{KL}}\) \cite{zou2022tbrats}), which uses a Dirichlet-based term to guide uncertainty estimation. The total loss is formulated as $L = \mathcal{L}_{\mathrm{seg}} + \mathcal{L}_{\mathrm{KL}}$. The segmentation loss is a weighted sum of Dice and weighted cross-entropy (WCE) losses, $\mathcal{L}_{\mathrm{seg}} = \gamma \mathcal{L}_{\mathrm{Dice}} + (1-\gamma) \mathcal{L}_{\mathrm{WCE}}$, with $\gamma$ empirically set to 0.6.

\section{Experiments and Results}
\subsection{Datasets and Implementation}
\textbf{Dataset.}
We evaluated our method on three datasets. 
CCTA119 is our in-house dataset,  which includes 119 CCTA volumes from a Grade III Level A medical institution, with a resolution of \( a \times a \times b \, \text{mm}^3 \) (\( a \in [0.28, 0.41] \), \( b \in [0.5, 1.0] \)) and a matrix size of \( 512 \times 512 \times N \) (\( N \in [155, 353] \)), annotated by three radiologists with at least 5 years of experience. 
The second is the MICCAI 2020 ASOCA challenge dataset \cite{gharleghi2023annotated}, containing 40 CCTA scans. The third, ICAS-100, is a subset of the ImageCAS dataset \cite{zeng2023imagecas}, consisting of 100 CCTA scans.

\textbf{Evaluation Metrics.} 
We evaluate our proposed method using two metrics: dice similarity coefficient (DSC) and average symmetric surface distance (ASSD).

\textbf{Implementation Details.}
We evaluated the proposed framework using five-fold cross-validation on the CCTA119 dataset (95 training, 24 testing subjects), ASOCA (32 training, 8 testing subjects) and ICAS-100 (80 training, 20 testing subjects).
All experiments were conducted using the PyTorch framework on NVIDIA 3090 GPUs. The networks were trained with the Adam optimizer, an initial learning rate of \(1 \times 10^{-4}\), 600 epochs, and a batch size of 2. During training, sub-volumes of size \(160 \times 160 \times 128\) were randomly cropped from the full volumes. In the testing phase, a sliding window approach with the same sub-volume size was used, moving in steps of half the window size to cover the entire volume. 
\begin{table}[!t]\footnotesize
	\caption{\label{tab-ccta}Comparison of different methods for coronary artery segmentation.
	}
	\setlength{\tabcolsep}{0.25mm}{
		\begin{tabular}{c|cc|cc|cc}
			\hline
			\multirow{2}{*}{Method}                                                                 &\multicolumn{2}{c|}{CCTA119}   
			&\multicolumn{2}{c|}{ASOCA}                                                              &\multicolumn{2}{c}{ICAS-100}                                                                                                                                                                                                                 \\ \cline{2-7} 
		& {{DSC $\%$$\uparrow$}}& {{ASSD $mm$$\downarrow$}}     	& {{DSC $\%$$\uparrow$}}  & {{ASSD $mm$$\downarrow$}}  & {{DSC  $\%$$\uparrow$}}   & {{ASSD $mm$$\downarrow$}}    \\ \hline
		3D-UNet\cite{cciccek20163d}   & {81.65±0.73}  & {1.62±0.12} & {84.78±0.85} & {1.41±0.13} & {76.17±1.04}	&1.57±0.16  \\ 
		S$^2$CA-Net\cite{zhou2024shape}& {84.01±0.68}   & {1.12±0.10}     & {87.11±0.79} & {1.16±0.13}   &78.88±0.87	&1.28±0.14 \\ 
		I$^2$U-Net \cite{dai2024i2u} & {84.56±0.64}     & {0.96±0.09}      & {87.47±0.72} & {0.85±0.12} &79.25±0.85	&1.12±0.12	\\ 
		UNETR\cite{hatamizadeh2022unetr}   & {83.22±0.57}  & {1.56±0.11} & {86.54±0.82}  & {1.21±0.13}  &78.64±0.83	&1.37±0.14  \\ 
		TransUNet\cite{chen2024transunet}& {83.53±0.62} & {1.52±0.09} & {87.38±0.69}  & {1.18±0.11}    & {78.87±0.82} &{1.32±0.13}   \\ 
		nnFormer\cite{zhou2023nnformer}& {84.61±0.56}    & {1.38±0.11} & {87.46±0.67}  & {1.03±0.10} &79.22±0.79	&1.15±0.14 \\ 
		CS$^2$Net\cite{mou2021cs2}& {84.34±0.61}      & {1.04±0.10}      & {87.01±0.62}& {0.88±0.12}    &79.07±0.71	&1.09±0.12\\ 
		\makecell[c]{3D-FFR\\-UNet\cite{song2022automatic}} &{84.58±0.55}      & {0.91±0.09}&87.12±0.64      & {0.82±0.10}    &78.94±0.83	&1.05±0.12      \\ 
		VSNet \cite{xu2025vsnet} & {85.07±0.51}      & {0.87±0.09}      & {88.04±0.63}& {0.79±0.11}    & {79.19±0.86} & {1.02±0.13}\\ \hline
		Ours& {\xblue{87.31±0.42}} & {\xblue{0.71±0.08}}      & \xblue{90.15±0.57}  & {\xblue{0.66±0.09}}   &\xblue{82.15±0.73} &\xblue{0.86±0.12} \\ \hline   
	\end{tabular}}
\end{table}	
\begin{figure*}[!ht]
	\centering
	\includegraphics[width=0.9999\linewidth]{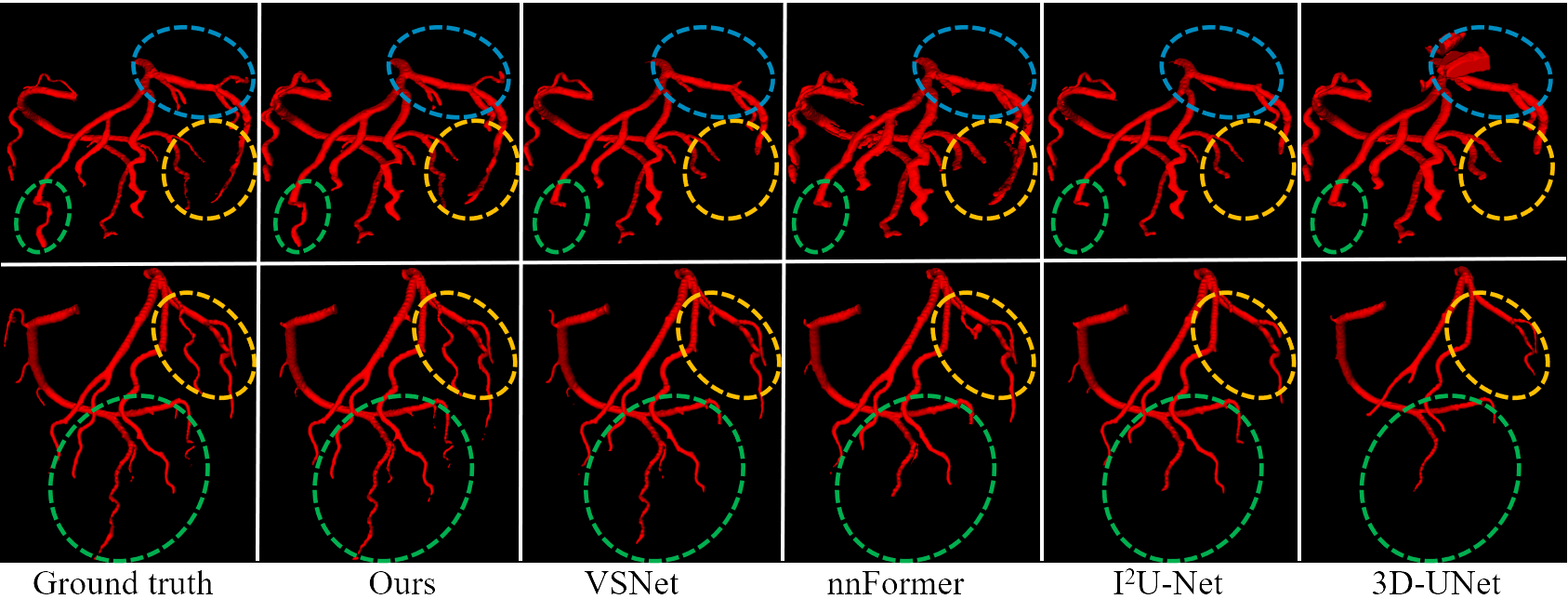}
	\caption{
		Visual results. The cyan, yellow and green dashed circles highlight the regions for better visual comparison.
	}
	\label{fig_ccta}
	 \vspace{-3ex}
\end{figure*}	
\subsection{Comparison with State-of-the-Art Methods}
We compared our method against nine state-of-the-art approaches, including CNN-based methods (3D-UNet \cite{cciccek20163d}, S$^2$CA-Net \cite{zhou2024shape}, I$^2$U-Net \cite{dai2024i2u}), transformer-based methods (UNETR \cite{hatamizadeh2022unetr}, TransUNet \cite{chen2024transunet}, nnFormer \cite{zhou2023nnformer}), and vessel segmentation methods (CS$^2$Net \cite{mou2021cs2}, 3D-FFR-UNet \cite{dong2022novel}, VSNet \cite{xu2025vsnet}). 
All comparisons used publicly available codes for fairness. Quantitative and qualitative results are shown in Table \ref{tab-ccta} and Fig. \ref{fig_ccta}, respectively, while Table \ref{cross-validation} presents cross-validation results, demonstrating the generalization capability of the proposed method.
%

\textbf{Quantitative Results.} As shown in Table \ref{tab-ccta}, we conducted extensive comparative experiments on CCTA119, ASOCA, and ICAS-100.  On \textit{CCTA119}, our method consistently outperforms all comparison methods, achieving a 5.66$\%$ higher DSC and a 0.91mm lower ASSD than 3D-UNet, as well as a 2.75$\%$ improvement in DSC over I$^2$U-Net, the best-performing CNN model. Furthermore, it surpasses the strongest transformer-based and vessel segmentation methods, with DSC gains of 2.70$\%$ over nnFormer and 2.24$\%$ over VSNet. On \textit{ASOCA}, our method also achieves the best performance among all compared methods, outperforming VSNet by 2.11$\%$ in DSC and 0.13 mm in ASSD. On \textit{ICAS-100}, our method performs consistently best among all comparisons, despite potential labeling errors in the dataset leading to relatively low overall metrics.

\textbf{Qualitative Results.} Fig. \ref{fig_ccta} provides a qualitative comparison, highlighting that comparison methods exhibit over-segmentation, under-segmentation, or both, leading to suboptimal performance. In contrast, our method closely aligns with ground truth, particularly in complex vascular structures, further demonstrating its effectiveness.

\textbf{Cross-validation Results.} 
As shown in Table \ref{cross-validation}, our method demonstrates superior performance in cross-validation. When trained on the CCTA119 dataset and tested on the ASOCA dataset, it achieves a DSC of 85.26\%, surpassing 3D-UNet by 6.12\% (79.14\%) and VSNet by 2.59\% (82.67\%). Moreover, our method achieves an ASSD of 0.83 mm, outperforming 3D-UNet by 0.81 mm (1.64 mm) and VSNet by 0.19 mm (1.02 mm). Consistent results on the \textit{CCTA119$\rightarrow$ICAS-100} setup further validate the method's strong generalization capability.
\subsection{Ablation Study}
We conduct an ablation study to evaluate the contributions of the key components in our proposed method: the Enhanced-ViT (ViT encoder enhanced by activating the final two ViT blocks and integrating the AGE module), the CVF module, and the EUR module. Starting from the baseline encoder-decoder network (Net1) \cite{zhang2018road},
we incrementally integrate the following components: Net2 adds Enhanced-ViT with sum-based fusion, Net3 replaces the sum fusion with CVF, Net4 extends Net1 by adding EUR, and Ours combines all components.

As shown in Table \ref{abs}, Net2 achieves a 1.25$\%$ improvement in DSC over Net1. Net3 further improves DSC by 1.21$\%$ over Net2, while Net4  shows a 1.19$\%$ gain over Net1. By integrating Enhanced-ViT, CVF, and EUR, Ours achieves a significant DSC improvement of 4.39$\%$ over Net1, demonstrating the effectiveness of each component and their synergistic effects in enhancing the framework.  
\begin{table}[t]\footnotesize
	\centering
	\begin{minipage}{0.64\linewidth}
		\centering
		\caption{\label{cross-validation} Cross-validation results: Model trained on CCTA119 and tested on ASOCA and ICAS-100.}
		\setlength{\tabcolsep}{0.01mm}{
		\begin{tabular}{l|cc|cc}
			\hline
				\multirow{2}{*}{Method}                                                                 &\multicolumn{2}{c|}{CCTA119$\rightarrow$ASOCA}   
			&\multicolumn{2}{c}{CCTA119$\rightarrow$ICAS-100} \\ \cline{2-5} 
			& {{DSC $\%$$\uparrow$}}& {{ASSD $mm$$\downarrow$}}     	& {{DSC $\%$$\uparrow$}}  & {{ASSD $mm$$\downarrow$}}     \\\hline 
			3D-UNet    & {79.14±0.81}	&1.64±0.17  &73.55±1.06 &1.62±0.18\\
			I$^2$U-Net         &82.36±0.75	&1.10±0.11  &75.76±0.87  &1.33±0.14\\
			nnFormer      &82.39±0.69	&1.09±0.12  &75.62±0.91	 &1.35±0.15\\
			VSNet       & {82.67±0.74} & {1.02±0.13} &75.87±0.94	 &1.28±0.14\\\hline
			Ours   	&\xblue{85.26±0.61} &\xblue{0.83±0.11} 	&{\xblue{78.74±0.82}} & {\xblue{1.05±0.13}} \\
			\hline        
		\end{tabular}}
	
	\end{minipage}
	\hfill
	\begin{minipage}{0.34\linewidth}
		\centering
		\caption{\label{abs} Ablation studies of our method on the CCTA119 dataset.}
		\setlength{\tabcolsep}{0.05mm}{
			\begin{tabular}{l|cc}
			\hline
			\multirow{1}{*}{Net} & \multirow{1}{*}{DSC $\%$$\uparrow$} &  \multirow{1}{*}{ASSD $mm$$\downarrow$}\\ 
			\hline
			Net1     &82.92±0.53 &1.39±0.09\\
			Net2          &84.17±0.46  &1.15±0.08\\
			Net3       &85.38±0.44	 &0.95±0.10\\
			Net4       &84.11±0.47	 &1.12±0.09\\\hline
			Ours   	&{\xblue{87.31±0.42}} & {\xblue{0.71±0.08}} \\
			\hline        
	\end{tabular}}
		
	\end{minipage}
\end{table}
\section{Conclusion}
In this study, we propose a novel segmentation framework that leverages the power of the VFM through a parallel encoding architecture for accurate coronary artery segmentation. The framework incorporates: 1) a ViT encoder to capture global high-level features and a CNN encoder to extract local low-level details, 2) a CVF module for adaptive feature integration via latent distribution modeling and variational attention, and 3) an EUR module to quantify uncertainty and refine segmentation by incorporating multi-scale feature information and attention mechanisms. Extensive experiments on in-house and public datasets demonstrate that our method outperforms state-of-the-art approaches, showcasing its effectiveness, robustness, and strong generalization capability. These results highlight its potential for advancing CAD diagnosis and clinical decision-making. 

\section*{Acknowledgements}
This research is supported by the National Natural Science Foundation of China (Grant Nos. 62301413, 82302309, 62371270, 12326617, and 12026609) and Natural Science Basic Research Program of Shaanxi (Grant No. 2025JC-YBQN-1134). 

\section*{Disclosure of Interests}
This study and its authors have no competing interests.

\bibliographystyle{splncs04}
\bibliography{refs}
\end{document}